\documentstyle[aps,preprint]{revtex}

\def\[{\left\lbrack}
\def\]{\right\rbrack}

\def\({\left(}
\def\){\right)}
\def\ih{\'\i}

\begin{document}

\title{Gauging by St\"uckelberg  field-shifting symmetry}

\author{Jorge Ananias Neto\thanks{jorge@fisica.ufjf.br}}
\address{Departamento de F\ih sica, ICE,\\ Universidade Federal de Juiz de Fora, 36036-330, Juiz de Fora, MG, Brazil }

\maketitle

\begin{abstract}
We embed second class constrained systems by a formalism that combines concepts of the BFFT method and the unfixing gauge formalism. As a result, we obtain a gauge-invariant system where the introduction of  the Wess-Zumino (WZ) field is essential. 
The initial phase-space variables are gauging with the introduction of the WZ field, a procedure that resembles the 
St\"uckelberg field-shifting formalism. In some cases, it is possible to eliminate the WZ field and, therefore, obtain an invariant system written only as a function of the original phase-space variables. We apply this formalism to important physical models:  the reduced-SU(2) Skyrme model and the two dimensional chiral bosons field theory. In  these systems, the gauge-invariant Hamiltonians are derived in a very simple way when compared with other usual formalisms.
\end{abstract}
\vskip 1 cm
PACS numbers: 11.10.Ef; $\,$11.15.-q $\\$
\newpage

\setlength{\baselineskip}{20pt}  
\renewcommand{\baselinestretch}{2}

\section{Introduction}

It is well known that first class theories or gauge theories, due to the presence of symmetries, describe in a more general way the physical properties of  constrained dynamical systems. These symmetries, in quantum field theory, can be used to deal with important questions as renormalisability and unitarity.  Almost all known fundamental interactions are described by first class theories. 

The designation  ``first class"  belongs to the Dirac's conventional formalism\cite{Dirac} where constrained systems are classified as first class theories and second class ones. First class constraints are considered to be the gauge-symmetry generators while the second class constraints are the reducers of the physical degrees of freedom. Consequently, in principle, there are no symmetries present in the dynamics of the second class systems.

It is possible to convert second class systems into first class ones.
The gauge-invariant systems must describe the same physical properties of the original second class models. Usually, there are two different approaches. One is the traditional formalism proposed by Faddeev and Shatashvili\cite{FS} and improved by Batalin and Tyutin\cite{BFFT,embed,JW}. In this approach,   WZ variables are added to the original system, equal in number to the number of second class constraints.
All the second class constraints and the second class Hamiltonian are changed in order to satisfy a first class algebra.
The second approach is the unfixing gauge formalism which has an opposite concept of the BFFT formalism. It was proposed by Mitra and Rajaraman\cite{MR} and improved by Vytheeswaran\cite{Vyt}. In this formalism, half of the second class constraints are considered to be the gauge-symmetry generators while the remaining ones are regarded to be the gauge-fixing terms. The second class Hamiltonian must be modified in order to satisfy a first class algebra with the constraints initially chosen to be the gauge-symmetry generators. This approach has an interesting property that does not extend the phase-space with extra variables.

However, in the chiral bosons field theory\cite{siegel,FJ,Ghosh} there is only one constraint. Due to this property, it is not possible to apply the unfixing gauge formalism in this system because this constraint satisfies a second class algebra. Motivated by this difficult, we propose a new scheme of first class conversion formalism that combines concepts of the BFFT method and the unfixing gauge formalism. Initially, we have proposed in Ref. \cite{OUR} a new first class conversion formalism which the gauge-invariant Hamiltonian must be directly obtained in order to be invariant by gauge-symmetry transformations. Now, in this paper the WZ fields are introduced with the objective to construct a gauge-invariant generator and a gauge-invariant phase-space variables.  Any function of these first class variables will be invariant by gauge transformation. 
This procedure resembles the St\"uckelberg field-shifting formalism\cite{stuckelberg,KHP,HKP} and, as we will see, simplifies, considerably, the algebraic calculations. As many important constrained systems have only two second class constraints, then, in principle, we describe the formalism only for systems with two second class constraints without any loss of generality.
It is clear that we are free to choose the second class constraint that will be selected to construct the gauge-invariant generator. The other second class constraint will be discarded. As an important result, we obtain a gauge-invariant version of the chiral bosons field theory extended with the WZ field. In some cases, it is possible to eliminate the WZ field and to derive a gauge-invariant system written only in terms of the original phase-space variables, a result that recovers the main idea of the unfixing gauge formalism. 

In order to clarify the exposition of the subject, this paper is organized as follows:  In Sec. II, we present the formalism in detail. In Sec. III, we apply this formalism to the collective coordinates expansion of the SU(2) Skyrme model\cite{Skyrme,ANW} and the two dimensional chiral bosons field theory.  These two physical systems are important nontrivial examples of the second class constrained systems.
The Skyrme model is a non-linear effective field theory which describes hadrons physics. In the chiral bosons field theory, the introduction of the WZ auxiliary field is essential to derive a gauge-invariant version.  In Sec. IV, we make our concluding remarks.

\section{The description of the formalism}

Consider a system in which the dynamics is governed by the Lagrangian ${\cal L}(q_i,\dot q_i)$ being $i=1,\dots,N$. The canonical Hamiltonian is obtained by performing the Legendre transformation, $H_c=p_i\dot{q}_i-{\cal L}$. Through the iterative Dirac's procedure in which states that the constraints have no time evolution, we determine the set of the two second class constraints written as

\begin{equation}
\label{cons}
T_a(q_i,p_i)\approx 0  \;\;{\text with \,\,}\,\, 
a=1,2\; .
\end{equation}

The formalism begins by constructing the symmetry generator as

\begin{eqnarray}
\label{fgen}
\tilde{T}=T_a+T_\theta,
\end{eqnarray}
where now $T_a$ is the second class constraint chosen to forge the symmetry generator and $T_\theta$ is a function of the WZ variables $(\theta,\pi_\theta)$. Further, $\tilde T$ must satisfy a first class Abelian algebra

\begin{eqnarray}
\{\tilde T, \tilde T \}=0.
\end{eqnarray}

All first class conversion formalisms, in principle, have some ambiguities\cite{BFFT,embed,JW,Vyt} and in our formalism, due to the arbitrariness of the algebraic form of $T_\theta$,  this situation is not different.  However,  we are free to make a convenient choice for $T_\theta$ in order to simplify the possible algebraic calculations or to exploit some new physical properties of the system. 

Representing the original phase-space variables as

\begin{equation}
\label{ft}
F=(q_i,p_i),
\end{equation}
thus our strategy is to construct a gauge-invariant function $\;\tilde A\;$ from the second class function $\;A \;$ by gauging the original phase-space variables. Denoting the first class variables by
 
\begin{equation}
\label{ftf}
\tilde F=(\tilde{q}_i,\tilde{p}_i),
\end{equation}
then we have the variational condition

\begin{equation}
\label{var}
\delta{\tilde F}=\{\tilde F,\tilde{T}\}=0,
\end{equation}
where $\tilde T$ is the symmetry generator defined in Eq.(\ref{fgen}). Any function of $\tilde F$ will be gauge-invariant since\cite{BB}

\begin{eqnarray}
\label{prop}
\{\tilde{A}(\tilde F),\tilde{T}\}=\{\tilde F,\tilde T\}
{\partial\tilde A\over \partial\tilde F}=0,
\end{eqnarray}
where 

\begin{equation}
\{\tilde F,\tilde T\}{\partial\tilde A\over \partial\tilde F}\equiv \{{\tilde q}_i,\tilde T\}{\partial\tilde A\over \partial{\tilde q}_i}+
\{{\tilde p}_i,\tilde T\}{\partial\tilde A\over \partial{\tilde p}_i}\;.
\end{equation}
Consequently, we can obtain a gauge-invariant function from the replacement of

\begin{eqnarray}
A(F)\rightarrow A(\tilde F)=\tilde A (\tilde F).
\end{eqnarray}

The gauge-invariant phase-space variables 
$\,\tilde F\,$ are built by adding an arbitrary function $G(F,\theta)$ to the original phase-space variables, namely

\begin{equation}
\label{FF}
\tilde F=F+G(F,\theta), 
\end{equation}
with the following boundary condition

\begin{equation}
\label{bondary}
G(F,\theta=0) = 0.
\end{equation}
Expanding the arbitrary function $G(F,\theta)$ 
in powers of $\;\theta$

\begin{equation}
\label{2060}
G(F,\theta)=G^1(F)\,\theta^1+G^2(F)\,\theta^2+\dots +G^n(F)\,\theta^n=\sum_{n=1}^\infty G^n(F)\,\theta^n\;,
\end{equation}
and imposing the variational condition, Eq.(\ref{var}), the corrections terms $G^n(F)$ and, consequently, the arbitrary function $G(F,\theta)$ 
can be completely determined. The general equation 
for $G^n (F)$ is

\begin{equation}
\label{variational}
\delta \tilde F = \delta F + \sum_{n=1}^\infty \left(\delta G^n\theta^n + n G^n\theta^{n-1}\delta\theta\right) = 0,
\end{equation}
where

\begin{eqnarray}
\delta F &=& \varepsilon\{F,{\tilde T}\} = \varepsilon\{F, T_a\},\nonumber\\
\delta G &=&\varepsilon\{G,{\tilde T}\}= \varepsilon\{G, T_a\}, \\
\delta\theta &=& \varepsilon\{\theta,{\tilde T}\}=\varepsilon\{\theta,T_\theta\}.\nonumber
\end{eqnarray}
Then, for the linear correction term ($n=1$), we have 

\begin{eqnarray}
\label{rdeltap1}
\delta  F +G^1\delta\theta &=& 0,\nonumber\\
G^1&=&-\delta F/\delta\theta.
\end{eqnarray}
For the quadratic correction term ($n=2$), we get 

\begin{eqnarray}
\label{rdeltap2}
\delta  G^1 + 2 G^2 \delta\theta & =& 0,\nonumber\\
G^2&=&-{1\over 2}\delta G^1/\delta\theta.
\end{eqnarray}
For $n\geq 2 $, the general relation is 

\begin{eqnarray}
\label{rdeltapn}
\delta G^n+(n+1) G^{n+1}\delta\theta &=& 0,\nonumber\\
G^{(n+1)}&=&-{1\over(n+1)}\delta G^n/\delta\theta.
\end{eqnarray}
Note that, in our formalism,  the recursion relations(\ref{rdeltap1}), (\ref{rdeltap2}) and (\ref{rdeltapn})  presuppose that the transformation for $\delta\theta$ must be linear, i.e., it is independent of $\theta$, since powers of $\theta$ are being compared.
Using again the relations (\ref{rdeltap1}), (\ref{rdeltap2}) and (\ref{rdeltapn}) we obtain the series

\begin{eqnarray}
\label{series}
\nonumber\\
\tilde F=F-{\theta\over\delta\theta}\delta F
+{1\over 2!}{\theta^2\over {(\delta\theta})^2}\delta\delta F-{1\over 3!}{\theta^3\over {(\delta\theta})^3}\delta\delta\delta F+\dots \;.
\end{eqnarray}
The expression (\ref{series}) can be elegantly written in terms of a projection operator on $ F $

\begin{eqnarray}
\label{proj}
\tilde F=e^{-{\theta\over\delta\theta}\delta}:F,
\end{eqnarray}
or
\begin{eqnarray}
\tilde F=e^{-\theta\,\hat\xi}: F\;,
\end{eqnarray}
where the operation $\;\hat{\xi}\,F$ is defined as 
$\;\hat{\xi}\,F\equiv{\{ F,T_a\}\over
\{\theta,T_\theta\}}\;.$

\vskip 1 cm

In order to eliminate the WZ auxiliary field we must find a representation for the WZ variable written only in terms of the original phase space variables $ F $, i.e.,\break $\theta=f(F)$. The algebraic form of this function is obtained imposing that it has the same infinitesimal gauge transformation displayed by  $\theta$, namely

\begin{equation}
\label{rep}
\delta\theta=\delta f(F).
\end{equation}
Thus, it is possible to derive a gauge-invariant Hamiltonian, $\tilde H$, written only as a function of the original phase space variables $ F $ satisfying the first class algebra

\begin{equation}
\label{fca}
\{ \tilde{H}, T_a \} =0,
\end{equation}
where $T_a$ is the second class constraint initially chosen to forge the first class constraint that now becomes the gauge-symmetry generator.

\section{Applications of the formalism}

\subsection{The reduced-SU(2) Skyrme model}

The Skyrme model describes baryons and their interactions through soliton solutions of the non-linear sigma model-type Lagrangian given by

\begin{eqnarray}
\label{Sky}
L = \int \, d^3x \[ {f_\pi^2\over 4} Tr\, (\partial_\mu U \partial^u U^+) + {1 \over 32 e^2 } Tr [ U^+\partial_\mu U,
U^+\partial_\nu U ] ^2 \],
\end{eqnarray}
where $f_\pi$ is the pion decay constant, $e$ is a dimensionless parameter and $U$ is a SU(2) matrix. 
Performing the collective semi-classical 
expansion\cite{ANW} just substituting $U(r,t)$ by $U(r,t)=A(t)U_0(r)A^+(t)$ in Eq. (\ref{Sky}), being $A$ a SU(2) matrix, we obtain 

\begin{equation}
\label{Lag}
L = - M + \lambda \,Tr [ \partial_0 A\partial_0 A^{-1} ],
\end{equation}
where $M$ is the soliton mass and $\lambda$ is the moment of inertia\cite{ANW}. The SU(2) matrix $A$ can be written as $A=a_0 +i a\cdot \tau$, where $\tau_i$ are the Pauli matrices, and satisfies the spherical constraint relation

\begin{equation}
\label{pri}
T_1 = a_i a_i - 1 \approx 0, \,\,\,\, i=0,1,2,3.
\end{equation}

\noindent Then, the Lagrangian (\ref{Lag}) can be read as a function of $a_i$ as

\begin{equation}
\label{cca}
L = -M + 2\lambda \dot{a}_i\dot{a}_i.
\end{equation}
Calculating the canonical momenta

\begin{equation}
\label{cm}
\pi_i = {\partial L \over \partial \dot{a}_i} = 4 \lambda \dot{a}_i,
\end{equation}
and using the Legendre transformation, the canonical Hamiltonian is computed as

\begin{eqnarray}
\label{chr}
H_c=\pi_i \dot a_i-L &=&
 M+2 \lambda \dot a_i\dot a_i \nonumber \\
&=&M+{1\over 8 \lambda } \sum_{i=0}^3 \pi_i\pi_i.
\end{eqnarray}
A typical polynomial wave function,
${1\over N(l)}(a_1 + i a_2)^l = |polynomial \rangle\, ,$ is an 
eigenvector of the Hamiltonian (\ref{chr}). This wave function is also eigenvector of the spin and isospin operators, written in \cite{ANW} as $ J_k={1\over 2}
( a_0 \pi_k -a_k \pi_0 - \epsilon_{klm} a_l \pi_m )$  and 
$ I_k={1\over 2 } ( a_k \pi_0 -a_0 \pi_k- \epsilon_{klm} a_l\pi_m ).$

From the temporal stability condition of the spherical constraint, Eq.(\ref{pri}), we get the secondary constraint

\begin{equation}
\label{T2}
T_2 = a_i\pi_i \approx 0 \,\,.
\end{equation}
We observe that no further constraints are generated via this iterative procedure. $T_1$ and $T_2$ are the second class constraints which the matrix elements of their Poisson brackets read as

\begin{equation}
\label{Pa}
\Delta_{\alpha \beta} = \{T_\alpha,T_\beta\} = -2 \epsilon_{\alpha \beta}
a_i a_i, \;\; \alpha,\beta = 1,2
\end{equation}
where $\epsilon_{\alpha \beta}$ is the antisymmetric tensor normalized as $\epsilon_{12} = -\epsilon^{12} = -1$.

In order to obtain a gauge-invariant SU(2) Skyrme model, the first step is to construct the extended generator of symmetry, which we choose as

\begin{equation}
\label{fcs}
\tilde{T}=T_1+\pi_\theta=a_ia_i-1+\pi_\theta.
\end{equation}
The infinitesimal gauge transformations generated by the symmetry generator ${\tilde T}$ are

\begin{eqnarray}
\label{IGTN}
\delta a_i &=& \varepsilon\{a_i,{\tilde T}\} = \varepsilon\{a_i, T_1\}=0,\nonumber\\
\delta \pi_i &=&\varepsilon\{\pi_i,{\tilde T}\}= \varepsilon\{\pi_i, T_1\}
=-2\varepsilon a_i,\\
\delta\theta &=& \varepsilon\{\theta,{\tilde T}\}=\varepsilon\{\theta,\pi_\theta\}=\varepsilon,\nonumber
\end{eqnarray}
where  $\varepsilon$ is an infinitesimal parameter. 
From the functional form of the second class Hamiltonian, Eq.(\ref{chr}), we see that the momentum $\pi_i$ is the only original phase-space variable that is necessary to shift in order to obtain a gauge-invariant Hamiltonian.  Then, the second step of the formalism is to construct the invariant momentum which read as

\begin{eqnarray}
\label{HF1}
\tilde{\pi}_i &=& \pi_i + G_i(a_i,\pi_i,\theta)\nonumber\\
&=& \pi_i+ G_i^1\,\theta + G_i^2\,\theta^2+ \dots +G_i^n\,\theta^n.
\end{eqnarray}
From the invariance condition $\delta{\tilde \pi}_i =0$  given in Eq.(\ref{variational}) and using the infinitesimal gauge transformations (\ref{IGTN}), we can compute all the correction terms $G^n$ given in 
Eq.(\ref{HF1}). For the linear correction term in order of $\theta$, Eq.(\ref{rdeltap1}), we get

\begin{eqnarray}
\label{IGT}
\delta \pi_i +  G_i^1 \delta\theta  & = &  0,\nonumber\\-2\varepsilon a_i +  \varepsilon G_i^1&=& 0,\\
 G_i^1 &=&  2a_i .\nonumber
\end{eqnarray}
For the quadratic term, we  obtain $ G_i^2= 0$, since $\delta G_i^1=\lbrace G_i^1,{\tilde T}\rbrace = 0$. Due to this,  all correction terms $ G_i^n$ with $n\geq 2$ are null. Therefore, the gauge-invariant momentum is

\begin{eqnarray}
\label{gim}
\tilde{\pi}_i = \pi_i+2a_i\theta,
\end{eqnarray}
where by using Eqs.(\ref{IGTN}), it is easy to show that, $\delta{\tilde\pi}_i=0$. The gauge-invariant Hamiltonian can be obtained in a very simple way as

\begin{eqnarray}
\label{gih}
\tilde{H}&=&{1\over 8\lambda}\tilde{\pi}_i\tilde{\pi}_i\nonumber\\
&=&{1\over 8\lambda}\pi_i\pi_i+{1\over 2\lambda}a_i\pi_i\theta
+{1\over 2\lambda}a_ia_i\theta^2.
\end{eqnarray}
This Hamiltonian, due to the relation in Eq.(\ref{prop}), satisfies the gauge-invariance property 

\begin{equation}
\{ \tilde{H},\tilde{T} \}=0. 
\end{equation}
In the gauge-invariant Hamiltonian, expressed in Eq. (\ref{gih}), if we fix the Wess-Zumino variable equal to zero, i.e., the unitary gauge, we recover the initial second class Skyrme model. 
\vskip .5 cm
From the infinitesimal transformation $\;{\delta\theta} = \varepsilon\;$, Eq.(\ref{IGTN}), we can choose a representation for $\theta$ as

\begin{eqnarray}
\label{fix}
\theta = f(a_i,\pi_i)=- {a_i\pi_i\over 2 a^2},
\end{eqnarray}
since $\;\delta f=\varepsilon\;$. Substituting the relation above in the Eq.(\ref{gim}), we get the invariant momentum written only in terms of the original phase-space variables, read as 

\begin{equation}
\label{IM}
{\tilde \pi}_i=\pi_i-a_i {a_j\pi_j\over a^2}.
\end{equation}
Consequently, from Eq.(\ref{gih}) we obtain the gauge-invariant Hamiltonian written only in terms of the original phase-space variables, given by 

\begin{eqnarray}
\label{HF3}
\tilde{H} &=& M + {1\over 8\lambda} \[ \pi_i\pi_i -  {(a_i\pi_i)^2\over a^2}\],\nonumber\\
&=& M + \frac {1}{8\lambda}\pi_i M^{ij}\pi_j ,
\end{eqnarray}
being the phase space metric $M^{ij}$ defined by

\begin{equation}
\label{metric}
M^{ij} = \delta^{ij} - {a^i a^j\over a^2}.
\end{equation}
The Hamiltonian (\ref{HF3}) is invariant under the infinitesimal gauge transformations, Eqs.(\ref{IGTN}), and due to this, the original second class constraint $\,T_1$ , Eq.(\ref{pri}), becomes the gauge symmetry generator. 

Here, we can observe  the auxiliary tool characteristic of the WZ variable $\theta$ because, at first, the WZ variable is introduced in the second class variables with the purpose to enforce the symmetries. Next, it is replaced by an adequate representation leading to reveal the hidden symmetry present in the original phase-space variables.

From the first class Hamiltonian, Eq.(\ref{HF3}), the gauge-invariant Skyrmion Lagrangian should be of the form $\,{\tilde L} \sim  \dot{a}_i  \,{(M^{ij})}^{-1}\, \dot{a}_j $.
Due to the fact that the matrix $\, M\,$, Eq.(\ref{metric}), is singular, then,
in principle, it is not possible to obtain the first class Skyrmion Lagrangian written only in terms of the original phase-space variables. For more details see Ref. \cite{TDL}.

Now, let us consider the Poisson brackets of the first class variables $\tilde a_i=a_i$ and $\tilde \pi_j= \pi_j-a_j {a_i\pi_i\over a^2}$. After some algebraic calculations, we have

\begin{eqnarray}
\label{PBF}
\{\tilde a_i,\tilde a_j\}&=&0,\nonumber\\
\{\tilde a_i,\tilde \pi_j\}&=&\delta_{ij} -{\tilde a_i\tilde a_j\over {\tilde a}^2},\\
\{\tilde\pi_i,\tilde \pi_j\}&=&{1\over {\tilde a}^2}\(\tilde a_j\tilde \pi_i - \tilde a_i\tilde\pi_j\).\nonumber
\end{eqnarray}
This result is the same obtained when we calculate the Dirac brackets between the original second class variables $a_i$ and $\pi_j$. This situation also occurs in the BFFT quantization of O(3) nonlinear sigma model\cite{HKP} where this result is described by using the following scheme

\begin{equation}
\{\tilde A,\tilde B \} = \{ A, B \}_{D (A_\rightarrow\tilde A ,B_\rightarrow\tilde B)}.
\end{equation}
Then, this result possibly indicates the equivalence between our formalism and the BFFT method.

The quantum equivalence of our first class system and the initial second class Skyrme model can be show by using  the Dirac's first class procedure.  The physical wave functions must be annihilated by the first class operator constraint, which reads as

\begin{eqnarray}
\label{qope}
T_1 |\psi \rangle_{phys} &=& 0.
\end{eqnarray}
The physical states that satisfy (\ref{qope}) are

\begin{eqnarray}
\label{physical}
| \psi \rangle_{phys} = {1\over V } \, 
\delta(a_i a_i-1)\, |polynomial\rangle.
\end{eqnarray}
where {\it V } is the normalization factor and $|polynomial \rangle ={1\over N(l)} (a_1+ i a_2)^l \,$. The corresponding quantum Hamiltonian is

\begin{eqnarray}
\label{echs1}
\tilde{H}= M+{1\over 8\lambda} \[ \pi_i\pi_i
-  {(a_i\pi_i)^2\over a_ja_j}\] .
\end{eqnarray}
The spectrum of the theory is determined by
taking the scalar product of the invariant Hamiltonian, $_{phys}\langle\psi|\tilde{H}|\psi \rangle_{phys}\,$, given by

\begin{eqnarray}
\label{mes1}
_{phys}\langle\psi|\tilde{H}|\psi \rangle_{phys}=\nonumber \\
\langle polynomial |\,\,  {1\over V^2}  \int da_i\,
\delta(a_i a_i - 1)\,
\tilde{H}\,
\delta(a_i a_i - 1)\,
| polynomial \rangle .
\end{eqnarray}
Integrating over $a_i$, we obtain 

\begin{eqnarray}
\label{mes2}
_{phys}\langle\psi| \tilde{H} | \psi \rangle_{phys}=\nonumber \\
\langle polynomial | M + {1\over 8\lambda}\[ \pi_i \pi_i 
- (a_i\pi_i)^2 \] | polynomial \rangle=\nonumber\\ \langle polynomial | M + {1\over 8\lambda} \[ p_i p_i \]| polynomial \rangle,
\end{eqnarray}
where $p_i\equiv (\delta_{ij}-a_i a_j)\pi_j$. As we can observe, the invariant Hamiltonian in Eq.(\ref{mes2})
presents ordering problems, and we solve this problem adopting the Weyl ordering prescription\cite{TDL2} where we construct the symmetrized expression for $p_i$ as

\begin{eqnarray}
\label{ordering}
[p_i]_{sym}&=&{1\over 2}\[ (\delta_{ij}-a_ia_j)\pi_j+\pi_j(\delta_{ij}-a_ia_j)\]\nonumber\\
&=& -i(\partial_i-a_ia_j\partial_j-{5\over 2}a_i),
\end{eqnarray}
where we have replaced $\pi_i$ by $-i\partial/\partial_i$. Substituting expression (\ref{ordering}) in (\ref{mes2}), we obtain

\begin{eqnarray}
\langle polynomial| \,[p_ip_i]_{sym}\,
| polynomial \rangle=\nonumber\\
\langle polynomial|\, M +{1\over 8\lambda}
\[ \partial_j\partial_j+ \(OpOp+2Op+{5\over 4} \)\]
| polynomial \rangle =\nonumber\\
\label{EL}
\nonumber\\
 M+{1\over 8\lambda} \[ l(l+2)+{5\over 4} \],
\end{eqnarray}
where the operator $Op$ is defined as $Op\equiv a_i\partial_i$.  Note that the eigenvalues of the operator $Op$ are defined by the following equation: $Op|polynomial \rangle= l \,|polynomial \rangle$. 
In Eq.(\ref{EL}),  the regularization of delta function squared  like $\delta^2 (a_i a_i - 1 )$ is performed by using the delta relation, $ 2\pi\delta(0)=\lim_{k\rightarrow 0}\int dx \,e^{ik\cdot x} = \int dx = L.$ Then, we use the parameter $L$ as the normalization factor. It is important to point out that the energy levels, formula (\ref{EL}), is the same obtained in a constrained second class treatment of the SU(2) Skyrme model\cite{JAN}. Thus, this result indicates that the field-shifting gauge-invariant formalism produces a correct result when compared with the original second class system.

\subsection{Chiral bosons field theory}

Chiral bosons field theory has received considerable attention. 
In spite of the apparent simplicity, this model can be relevant to the comprehension of superstrings, W gravities, and general two-dimensional field theories in the light cone. 

The two-dimensional Floreanini-Jackiw (FJ) chiral boson model has the dynamics governed by the following Lagrangian density\cite{FJ}

\begin{equation}
\label{CBTL}
{\cal{L}}= \dot{\phi} \phi^\prime-{\phi^\prime} ^2,
\end{equation}
where dots and primes represent derivatives with respect to time and space coordinates, respectively. The primary constraint is

\begin{equation}
\label{CBTC}
T(\phi,\pi) = \pi - \phi^\prime,
\end{equation}
and the canonical Hamiltonian is

\begin{equation}
\label{CBTH}
{\cal{H}}_c= {\phi^\prime}^2.
\end{equation}
The additional constraint called a secondary constraint can be generated by the Dirac's iterative procedure. However, in the chiral boson field theory, the primary constraint $T$ itself becomes a second class constraint which satisfies the following Poisson bracket relation

\begin{equation}
\label{CBC}
\lbrace T(x), T(y)\rbrace = - 2\delta'(x - y).
\end{equation}
Thus, in order to obtain a gauge-invariant chiral boson field theory, the first step is to construct a gauge-invariant generator $\tilde T$ from the second class constraint $T$, which we choose as

\begin{equation}
\label{FCBT}
\tilde{T} = \pi - \phi^\prime + \theta,
\end{equation}
where the WZ auxiliary field satisfies a non canonical Poisson bracket relation

\begin{equation}
\label{NC}
\lbrace \theta(x), \theta(y)\rbrace = 2\delta'(x - y).
\end{equation}
Combining Eqs.(\ref{CBC}) and (\ref{NC}), we have the first class Poisson bracket

\begin{equation}
\label{algeb}
\lbrace \tilde T(x), \tilde T(y)\rbrace = 0.
\end{equation}
The gauge infinitesimal transformations generated by $\tilde T$ are

\begin{eqnarray}
\label{DCBT}
\delta \phi(x) & = & \varepsilon\,\, \{ \phi(x),\tilde{T}(y)\} = \varepsilon \delta(x-y),\nonumber\\
\delta \theta(x)&=& \varepsilon \,\{  \theta(x), \tilde{T}(y) \} = 2\,\varepsilon\delta'(x-y).
\end{eqnarray}
The first class variable is built by adding an arbitrary function $G$ in the field $\phi^\prime$

\begin{eqnarray}
\tilde\phi' &=& \phi' +  G(\phi,\pi_\phi,\theta)\nonumber\\
&=& \phi'+ G^1\theta + G^2\theta^2 +\dots.
\end{eqnarray}
Following the prescription of our formalism, the correction terms $G^n$ are obtained by imposing the variational condition $\delta{\tilde {\phi}}' = 0$. Then, using the variational condition and the relations ({\ref{DCBT}), the linear correction term is obtained as  

\begin{eqnarray}
\label{G1}   
\delta \phi' + G^1 \delta\theta &=& 0,\nonumber\\
 \varepsilon\delta'(x-y) +  2 \varepsilon\delta'(x-y) G^1 &=&0,\nonumber\\
G^1&=& -{1\over 2}.
\end{eqnarray}
As the first correction term is a number, all correction terms $G^n$, for $n \geq 2$, are null. Therefore, the gauge-invariant field is

\begin{equation}
{\tilde\phi}'=\phi'-{1\over 2}\theta,
\end{equation}
where it is easy to show that, using Eqs.(\ref{DCBT}),
$\,\delta{\tilde\phi}'=0$. The gauge-invariant Hamiltonian density can be obtained in a very simple way as

\begin{eqnarray}
\label{FHCBT2}
\tilde{\cal H} &=& ({\tilde\phi})^{\prime 2}\nonumber\\
&=& (\phi' - \frac 12 \theta)^2,\nonumber\\
&=&{\phi^\prime}^2 - \phi'\theta + \frac 14 \theta^2,
\end{eqnarray}
where due to the property, Eq.(\ref{prop}), satisfies a first class algebra

\begin{equation}
\lbrace \tilde{\cal{H}}, \tilde T \rbrace = 0,
\end{equation}
with $\tilde T = \pi - \phi^\prime + \theta$.

The gauge-invariant Hamiltonian, Eq.(\ref{FHCBT2}), is the same obtained by Amorim and Barcelos in \cite{AB} via BFFT 
formalism\footnote{Here, it is appropriate to comment that chiral boson field theory is an example which the BFFT scheme does not necessarily involve fields canonically conjugated to the Wess-Zumino fields.}  with the advantage that we have used few algebraic steps. Then, this result also indicates the equivalence between our field-shifting gauge-invariant formalism and the BFFT first class conversion method. 

We can obtain the corresponding Lagrangian density by means of the constrained path integral formalism and the result is the same obtained in Ref.  \cite{AB}. It is opportune to comment that in the chiral bosons model, at first, is not possible to choose an adequate representation for the WZ field in terms of the original phase space variables. It occurs due to the singular property of the FJ chiral bosons model, whose constraint, Eq.(\ref{CBTC}), satisfies a second class algebra, given in Eq.(\ref{CBC}). Thus, it is necessary, in principle, to add  the WZ variable  in the derivation of the first class algebra, Eq.(\ref{algeb}).

\section{Conclusions}

In this article, we have proposed a first class conversion formalism that combines concepts of the BFFT method, the unfixing gauge formalism and the St\"uckelberg field-shifting scheme. From a second class constrained system with two second class constraints, we choose one constraint to forge, with the aid of  the WZ auxiliary variable, the gauge-symmetry generator. From a projection operator, Eq. (\ref{proj}), we construct first class variables. Consequently, any function of these first class variables will be gauge-invariant functions. This procedure, as we have observed in the Skyrme model and in the chiral bosons field theory,  certainly leads to considerable simplifications in the derivation of the first class functions.  In some cases, it is possible to obtain a first class Hamiltonian written only as a function of the original variables, an important result that recovers the original concept of the unfixing gauge formalism. It is clear that a procedure that verifies if the resulting first class theory reproduces the same equation of motion or the spectrum(at a quantum level) of the initial second class model must be evaluated at the end of the application of the formalism. 
Two subjects can be investigated as complementary studies to be developed in future papers. The first is the extension of our gauge-invariant conversion formalism for constrained systems with more than two second class constraints. The second is the possibility that the symmetry generator and the first class Hamiltonian satisfy now a non Abelian algebra. 

\section{ Acknowledgments}
The author would like to thank  A. G. Sim\~ao for critical reading, and C. Neves and W. Oliveira for valuable discussions. This work is supported in part by FAPEMIG, Brazilian Research Agency.

\end{document}